\documentclass[conference]{IEEEtran}
\IEEEoverridecommandlockouts
\usepackage{cite}
\usepackage{amsmath,amssymb,amsfonts}
\usepackage{algorithmic}
\usepackage{adjustbox}
\usepackage{graphicx}
\usepackage{textcomp}
\usepackage{xcolor}
\usepackage{multirow}
\usepackage{multicol}
\def\BibTeX{{\rm B\kern-.05em{\sc i\kern-.025em b}\kern-.08em
    T\kern-.1667em\lower.7ex\hbox{E}\kern-.125emX}}
\begin{document}

\title{Oldie is Goodie: Effective User Retention by In-game Promotion Event Analysis}

\author{\IEEEauthorblockN{Kyoung Ho Kim}
\IEEEauthorblockA{\textit{Graduate School of Information Security} \\
\textit{Korea University}\\
Seoul, Korea \\
quixote@korea.ac.kr}
\and
\IEEEauthorblockN{Huy Kang Kim}
\IEEEauthorblockA{\textit{Graduate School of Information Security} \\
\textit{Korea University}\\
Seoul, Korea \\
cenda@korea.ac.kr}
}

\maketitle

\begin{abstract}
For sustainable growth and profitability, online game companies are constantly carrying out various events to attract new game users, to maximize return users, and to minimize churn users in online games. Because minimizing churn users is the most cost-effective method, many pieces of research are being conducted on ways to predict and to prevent churns in advance. However, there is still little research on the validity of event effects. In this study, we investigate whether game events influence the user churn rate and confirm the difference in how game users respond to events by character level, item purchasing frequency and game-playing time band.
\end{abstract}


\section{Introduction}
Massively Multiplayer Online Role-Playing Game (MMORPG) is one of the major genres of online games. Users create their characters and catch monsters, hunt on scenarios designed by the game designer; they perform quests to strengthen items, thereby raising experience levels and thus raising the level. The game provides a variety of contents for frequent user interaction activities such as collecting and producing items, a battle between users (PvP), party play, exchange of items, etc.

Research on the acquisition and maintenance strategy of game users directly related to the profit of the game company has become the most crucial concern in the game industry. Besides, since the cost of acquiring new customers is higher than the cost of maintaining existing customers \cite{b1}, research on securing and managing existing customers has been actively studied in various fields in conjunction with the needs of companies \cite{b2, b3}. Users in the game can lose interest in the game over time and then eventually leave the game. These churn not only lead to a decline in revenues of game companies, but they also harm other users and can lead to their departure \cite{b4}. As a result, game companies try to extend the life cycle of game users by providing ongoing events and rewards.

In this study, we confirm whether churn rate decreases when game companies provide events and promotions through analyzing AION's actual dataset. AION is one of the largest MMORPGs in east Asia, developed and serviced by NCSOFT, one of the most famous game companies in Korea. Besides, we extract a few characteristics such as the user's character level, item purchasing frequency, and game-playing time band among the logs in which are recorded the user's various activities; and we confirm how the events have different effects per extracted characteristics.

\section{Contribution}
\begin{itemize}
\item This paper is the first case to focus on engineering analysis with a business perspective in online game research.
\item We discover the event is effective, especially for old users.
\item We discover that the event affects different influences according to user characteristics. Thus, game companies know which events are the most effective for maximizing their profits.
\end{itemize}

\section{Related works}
User churn issues directly related to the profitability of an online service company are the eternal assignment to be solved in all online services industry. Online games have also made many attempts to solve these problems \cite{b5, b6}. Borbora and Srivastava proposed a method of comparing and analyzing the user’s lifecycle using a distance-based classification system called wClusterDist as a method for distinguishing between churn users and users who continue to play game \cite{b7}, Kawale \textit{et al.} analyzed the relationship between social activity and user churn and discovered that social activities are influencing user churn from the game \cite{b4}. Seo \textit{et al.} proposed a churn prediction model by measuring social activities such as co-play between guild members \cite{b8}. Kim \textit{et al.} presented the study results that the difference in the churn prediction performance varies according to the observation period and the churn prediction period through actual data of three games. According to the results of these researches, it is crucial to set an observation period and the churn prediction period according to game characteristics\cite{b9}. Kim \textit{et al.} analyzed bad effect from bot users in game. They revealed that low-level users (e.g., newbie users) can be easily affected by malicious bot users' game plays \cite{b10}. 

These studies can be categorized into survey-based, Open API-based, and server-log based on the data collection method, as shown in Table \ref{tab1}. 

\begin{table}[htbp]
\caption{Data collection methods for user churn analysis and related games and research}
\begin{center}
\begin{tabular}{|c|c|c|}
\hline
\textbf{Category} & \textbf{Related online game} & \textbf{Reference} \\
\hline \hline  
Survey & World of Warcraft & \cite{b5} \\
\hline
Open API & BATTLEGROUNDS & \cite{b12} \\
\hline
Server-Log based & AION & \cite{b8, b10, b11, b13, b16}\\
\hline
\end{tabular}
\label{tab1}
\end{center}
\end{table}

The first method, the survey-based method, is used to survey game users to extract the characteristics that want to verify and whether the user leaves the games. Typically, there is research on World of Warcraft game users, and this research investigated how user’s churn rate varies according to game motivation and demographic information. The research results show that users with strong motivation tend to have low churn rates, while those with strong social motivation tend to have higher churn rates\cite{b5}. 

The second method, Open-API (Application Programming Interface) is used to collect data provided by game companies. A typical example is the BATTLEGROUND game. There is an advantage in that a lot of log data about a game can be obtained by only registering without having a direct relationship with a game company. 

Finally, this is a method of estimating churn through the measure of the user’s behavior and churn based on analysis of stored logs in the game server. This method can have the highest prediction performance as much as it can utilize the most information. Recently, several game companies have provided a sample log to host a contest for churn prediction analysis \cite{b12}. 

Moon \textit{et al.} developed a model which can quickly estimate a user's prediction with 90\% of accuracy (4 weeks earlier than real churn occurs) by analyzing social activities such as quest, party-play and trading \cite{b11}. 

Lee \textit{et al.} proposed a model that optimizes profit by predicting churn of loyal customers who provided a lot of profits for a long time in AION, it confirmed that this method is more cost-effective than performing the churn prediction from the entire users after applying to the actual game environment \cite{b14}. 

Oh \textit{et al.} proposed a churn prediction method based on various user types. They classified churn users into 5 types (light user, light-contents enjoying, heavy-contents enjoying, item making, and hunting with party-play type) by game users' play style. They found 98.3\% of churn users are in the `light user' type \cite{b16}.

\section{Methodology}
In this section, we define the churn criterion used to analyze the event effect and describe the classification method of the user type used in this study, the event reference date which is the basis of the event effect measurement, and the methodology for analyzing the churn change by user type according to an event.
\subsection{Defining churn criterion}
A reasonable churn criterion must be established for accurate churn user classification. Defining appropriate measures that do not overestimate or underestimated churn users is essential to create a practical abstraction classifier model \cite{b15}. In this study, we refer to the method used by Yang \textit{et al.} \cite{b17} and define ‘a user who has no login for N days’ as a churn user. In this experiment, users who are not login continuously from 2 days to 6 days after login based on the start date of analyzed data are defined as churn user. A churn rate is calculated as follows.

Churn rate = (the number of users who have not logged in consecutively for N days) / (the number of cumulative users logged in per day)

\subsection{Classification of user type}
AION provides various analysis features on game logs. Among them, we selected the character’s level, item purchase frequency, and game-playing time band as features which are the most frequently used in churn prediction research as shown in Table \ref{tab2}.

\begin{table}[htbp]
\caption{Feature selection for analyzing user type}
\begin{center}
\begin{tabular}{|p{2.4cm}||p{5.2cm}| }
\hline
\textbf{Feature} & \textbf{Description} \\
\hline \hline
Character’s Level & Highest level of character before and after the event \\
\hline
Item purchase frequency & Number of purchases in store \\
\hline
Game-playing time band & Time band with the largest number of activity logs for two weeks \\
\hline

\end{tabular}
\label{tab2}
\end{center}
\end{table}

The character’s level correlates with game-playing time and loyalty to the game. Well-designed events can satisfy engaged users’ satisfaction, which increases game-playing time. Item purchase frequency is the most critical indicator of the game company’s profits. In general, user’s satisfaction and commitment to the game can increase item purchase. To check the increased number of items purchased after the event is a routine process to measure the event’s success by game companies. Game-playing time band is selected to identify the targeting users of the promotion. For example, we can classify the target user’s group (i.e., hard-core user and light-user) by analyzing game-playing time.

\begin{table}[htbp]
\caption{Criteria for classification of user type}
\begin{center}
\begin{tabular}{|p{2cm}||p{1.7cm}|p{1.7cm}|p{1.7cm}| }
\hline
\textbf{Feature}&\multicolumn{3}{|c|}{\textbf{Classification Criteria}} \\
\cline{2-4} 
\hline \hline
Character’s Level & High (61-80) & Mid (31-60) & Low (1-30) \\
\hline
Item purchase frequency & High (Over 10 times) & Mid (6 to 10 times) & Low (0 to 5 times) \\
\hline
Game-playing time band & Dawn (0-6) & Day (6-19) & Night (19-24)\\
\hline
\end{tabular}
\label{tab3}
\end{center}
\end{table}

\subsection{Selection of event reference date}
In the AION game, several types of events are played in succession from two to four weeks. Besides, since more than 60 events are held every year, it is difficult to measure the effects of individual events because there are no independent events to be performed at a specific time.

In this study, the event effects were analyzed by comparing the data before and after on 3 February 2016, when four events occurred simultaneously with the mid-day of the experiment data as shown in Table \ref{tab4}. All event information is retrieved from AION’s official web site \cite{b18}.

\begin{figure*}[]
\includegraphics[width=\textwidth,height=6cm]{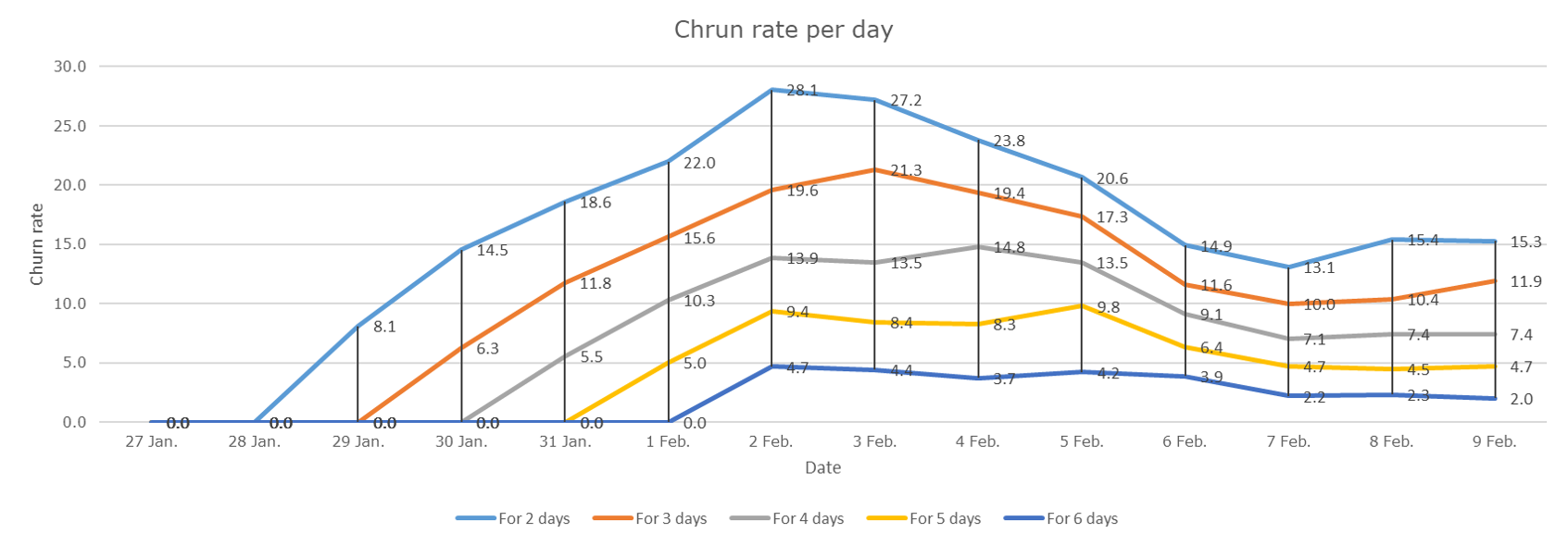}
  \caption{Churn rate per day during the observation period. It shows the churn rate gradually decreases from the event start time.}
  \label{fig2}
\end{figure*}

\begin{table}[htbp]
\caption{Event reference date using in the experiment}
\begin{center}
\begin{tabular}{|p{2.4cm}|p{1.7cm}|p{1.5cm}|p{1.5cm}| }
\hline 
\textbf{Event name}& \textbf{type} & \textbf{Start date of event} & \textbf{Event duration} \\
\hline \hline
Stigma          & Reinforcement  & 3 Feb, 2016 & 14 days \\
\hline
Sullenday      & Ballentine Day & 3 Feb, 2016  & 21 days \\
\hline
Strongest legions & Battle         & 3 Feb, 2016 & 24 days \\
\hline
Dr. Edishung      & Potion making  & 3 Feb, 2016 & 28 days \\
\hline
\end{tabular}
\label{tab4}
\end{center}
\end{table}

Each event's description is summarized as follows. 
\begin{itemize} 
\item \textbf{Stigma}: Stigma does not disappear even if a user fails to strengthen a stigma in the event period. Stigma usually vanishes into the air if a user fails to reinforce a stigma.

\item \textbf{Sullenday}: A user can purchase a Lucky Box with various items at a low price.

\item \textbf{Strongest legions}: Only level 46 or higher users can attend battles with legion team members, and they can get a reward as much as they win.

\item \textbf{Dr. Edishung}: If users purchase the ticket, they get essential items for level up and get a `40\% discount coupon' which can buy other tickets for 90 days or 900 hours.
\end{itemize} 

\subsection{Analysis of churn by user type according to an event}
\begin{figure}[h]
\centerline{\includegraphics[width=0.47\textwidth]{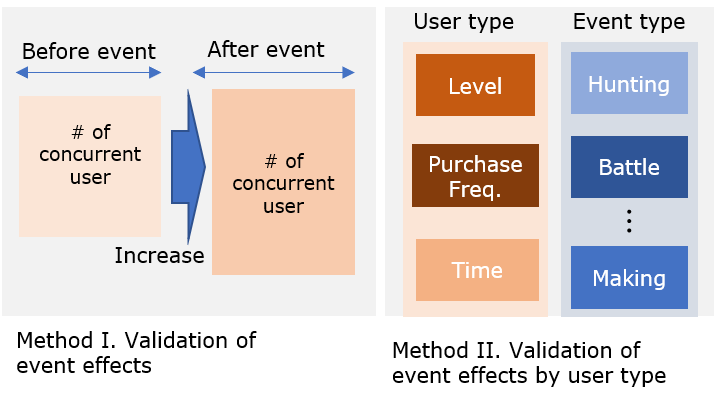}}
\caption{This image shows validation method of event effects. First, we verify the event effects by comparing numbers of login users before and after events. Second, we validate the event affects different influence per user type.}
\label{fig1}
\end{figure}

We propose two-step analysis with Method I and II as shown in Figure\ref{fig1}

\section{Experiment}
\subsection{Definition of churn criterion}
In this study, we used about 54 million actual log data from AION provided by NCSOFT, the most famous online game companies in Korea. This data contains about 20,000 real user game activity information for two weeks from 27 January 2016 to 9 February 2016. To calculate the percentage of churn users during the two weeks, we extracted all the information about the users who did not login for two to six days in two weeks. 

Figure \ref{fig2} shows the percentage of churn users within the observation period. For each event, it heats the peak value one day before the event date, and then churn rates gradually decrease from the event start time. This result indicates that the event is useful to prevent churn in online games. 

\begin{figure}[]
\centerline{\includegraphics[width=0.47\textwidth]{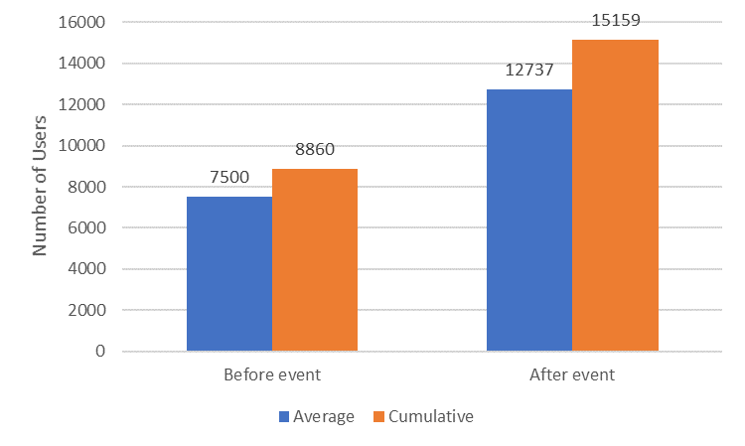}}
\caption{Comparison of the number of login users between before and after the event.}
\label{fig3}
\end{figure}

\subsection{Classification of user type}
Table \ref{tab5} summarizes the data classified according to the user type classification by analyzing the log data. First, AION’s maximum level is up to 80. The AION, which passed more than seven years since its first release, shows that more than 50 percent of users have a level of 61 or higher. Second, we confirmed that the number of items purchased per user is less than five times per week. Third, we know that about 64\% of gameplay users are playing games during the daytime.

\begin{table}[htbp]
\caption{This table shows user distribution by classes}
\begin{center}
\begin{tabular}{|p{2cm}||p{1.7cm}|p{1.7cm}|p{1.7cm}| }
\hline
\textbf{Feature}&\multicolumn{3}{|c|}{\textbf{User Distribution by Classes}} \\
\cline{2-4} 
\hline
Character’s Level & High 50\% & Mid 45\% & Low 5\% \\
\hline
Item purchase frequency & High 13.5\%  & Mid 14.5\% & Low 72\% \\
\hline
Game-playing time band & Dawn 18\% & Day 64\% & Night 18\%\\
\hline
\end{tabular}
\label{tab5}
\end{center}
\end{table}

\subsection{Comparison of the number of active users between before and after the event}
Figure \ref{fig3} shows the number of active users before and after the event. Before the event, the average number of active users per week is 7,500, and the number of cumulative users is 12,737. After the event, the average number of active users 8,860 and the number of cumulative users is 18,159, which is increased by 18\% and 19\% respectively.

This result shows that the event is an effective way to increase the number of active users. However, it is difficult to confirm whether the increased user is an existing user, a new user, or a returning user.

\subsection{Distribution of user-level before and after the event}
\begin{figure}[h]
\centerline{\includegraphics[width=0.47\textwidth]{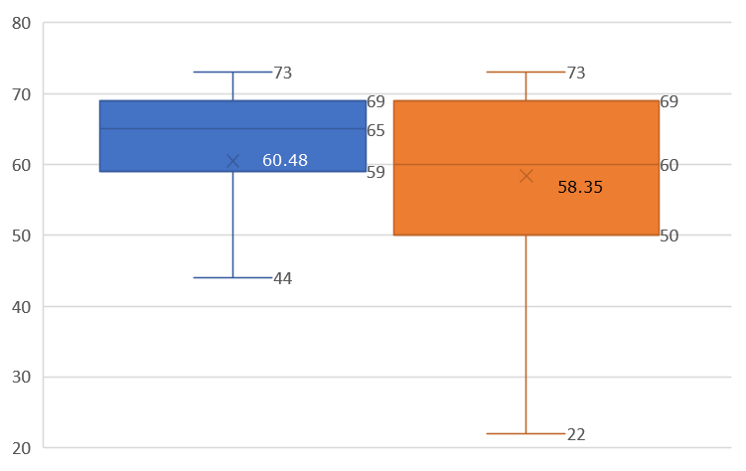}}
\caption{Distribution of user-level before and after the event; The blue bar shows the distribution of user-level before the event while the orange bar shows the distribution of user-level after the event. We know that the login of low-level users increases after starting the event.}
\label{fig4}
\end{figure}

Figure \ref{fig4} shows the level distribution of users before and after the event in quartiles. Before the event, the average level of login users was 60, but the average level after the event was lowered to 58, and the lowest level user was also lowered from 44 to 22 after the event. This graph shows that the event has a more significant impact on users at lower levels than average.

\subsection{Event effect analysis by character’s level class}
\begin{figure}[h]
\centerline{\includegraphics[width=0.47\textwidth]{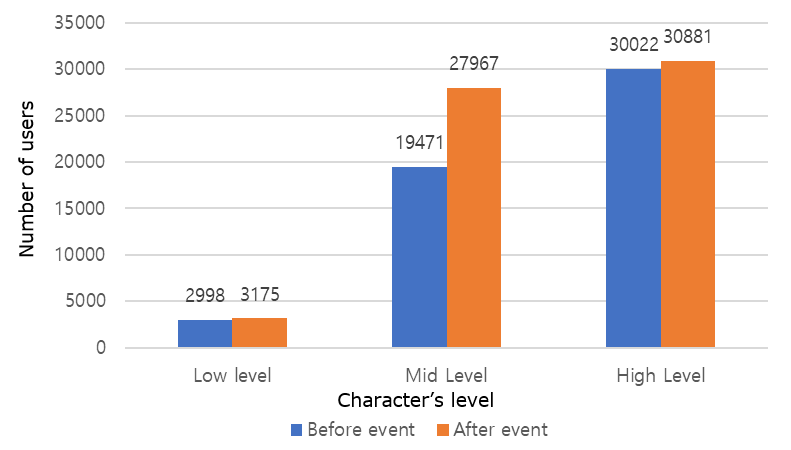}}
\caption{Event effect analysis by character’s level class; It shows mid-level user’s login is the most increased}
\label{fig5}
\end{figure}

Figure \ref{fig5} shows the analysis of the event effect by a character’s level class. At the event, the number of logins for the entire class level user has increased, especially at level 31 to 60, with middle-class user connections are growing to about 44\%.

\subsection{Event effect analysis by game-playing time bands}
Figure \ref{fig6} shows the effect of the time-based event on the game. It shows that dawn and daytime users increased while nighttime users decreased.

\subsection{Event effect analysis by item purchase frequency}
Figure \ref{fig7} shows the change in the item purchase frequency before and after the event. Contrary to expectations, purchase times after the event decrease by about 10\%, while the number of buyers increased by nearly 150\% due to event gifts. 
The number of users who buy game items more than ten times per week reduces significantly while the number of purchases under five times per week increases drastically.

If AION game logs are including the detailed purchase log, it is possible to check the effect of the event in monetary units. However, since the analyzed logs do not include the purchase item price, only the number of simple purchases and the number of purchase users were analyzed. 

\begin{figure}[h]
\centerline{\includegraphics[width=0.47\textwidth]{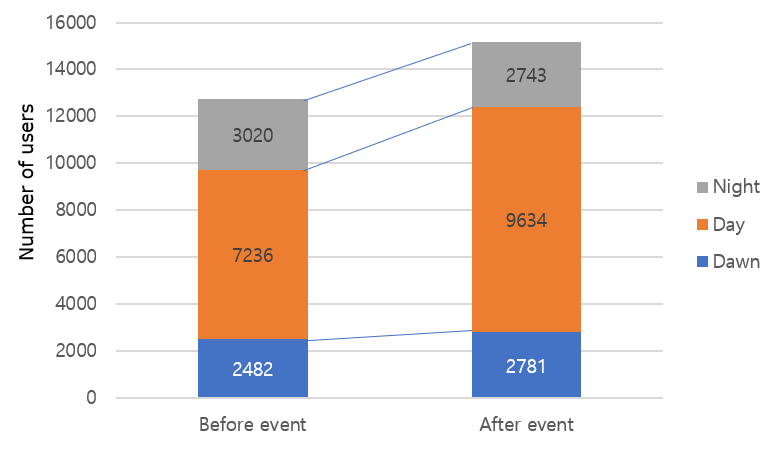}}
\caption{Event effect analysis by game-playing time band; It shows that the event influence on daytime users relatively.}
\label{fig6}
\end{figure}

\begin{figure}[h]
\centerline{\includegraphics[width=0.47\textwidth]{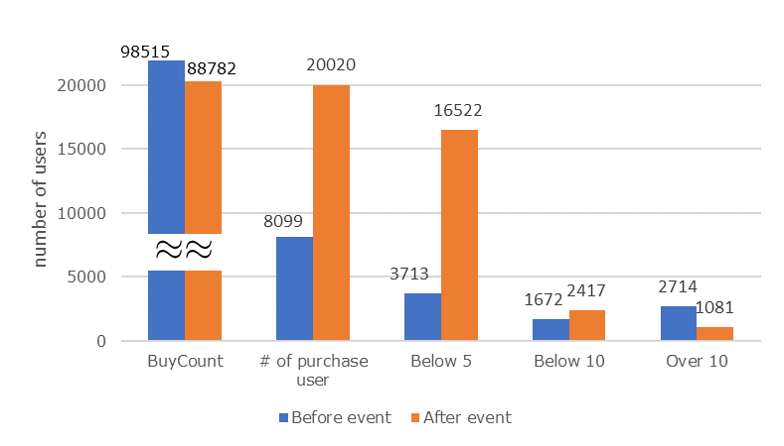}}
\caption{Purchase pattern change before and after the event; It shows the number of purchases over 10 times decrease.}
\label{fig7}
\end{figure}

Figure \ref{fig5} shows that mid-level user’s login is the most increased, Figure \ref{fig6} shows that the event influences on daytime users relatively, and Figure \ref{fig7} shows that the number of purchase decreases. If we analyze Figure \ref{fig5}, \ref{fig6} and \ref{fig7} collectively, we can find that users who play games during the daytime have a lot of mid-level and do not use the money for buying game items.
If game companies use the result of analyzing various event effects by the user's characteristics, the most profitable, effective marketing strategy can be formulated.

\subsection{Event effect analysis by game-playing time bands}
Figure \ref{fig6} shows the effect of the time-based event on the game. It shows that dawn and daytime users increased while nighttime users decreased.

\section{Conclusion and Future work}
In this study, we used the log data of AION, which is currently in service to check that game events are useful to churn prevention and how long time takes to generate effects to users in MMORPG. Besides, we confirmed that event effects are different depending on the character’s level, item purchase frequency and game-playing time band.

The most unusual aspect of the analysis is that the number of purchase activities in the game decreased even though the event has the effect of preventing the user churn and increasing the number of return users. This phenomenon means that additional analysis is needed internally for this fact in the game company to check whether the event results in substantial monetary gain.

We find the increase of new customers is minimal while the event increases the number of mid-level users who usually play in the daytime through the experiment. This finding shows that the event is more effective in preventing churn users rather than introducing new users. 

In this study, we find that the users who purchased more than ten times during the event period and who enjoyed playing in the nighttime decreases while the users who purchased less than five times and who enjoyed playing in the daytime increases. This analysis shows that the game companies need to focus on events for the daytime user to retain users while focusing on events for the night user to increase revenues. 

These results of these analyses are very instrumental and practical not only to game engineer but to game marketers. Thus, this study can be effectively used for marketing strategies and plans that can lead to an increase in return users as well as decreasing churn users. 

Furthermore, the event effect analysis proposed in this study will be able to have a significant synergy effect through the target marketing if it is analyzed together with the classification of the user characteristics type in the game conducted in other studies. 

In the future, we will verify the universality of the proposed method by testing with the extended period of AION dataset. Then, we will research the similarity and differences of event effects by analyzing other genre of games such as FPS (First Person Shooter games), arcade and so forth. Finally, we will study the how events in PC and mobile games have different effects.


\end{document}